\documentclass[aps,prl,twocolumn,superscriptaddress,showpacs]{revtex4}
\usepackage{graphicx}% Include figure files
\usepackage{dcolumn}% Align table columns on decimal point
\usepackage{bm}% bold math
\usepackage{epsfig}
\usepackage{longtable}
\def\Z{$\zeta$~}

\begin{document}

% Use the \preprint command to place your local institutional report
% number in the upper righthand corner of the title page in preprint mode.
% Multiple \preprint commands are allowed.
% Use the 'preprintnumbers' class option to override journal defaults
% to display numbers if necessary
%\preprint{}

%Title of paper
\title{Measurement of Transverse Single-Spin Asymmetries for Di-Jet
Production in Proton-Proton Collisions at $\sqrt{s} = 200$ GeV}

% repeat the \author .. \affiliation  etc. as needed
% \email, \thanks, \homepage, \altaffiliation all apply to the current
% author. Explanatory text should go in the []'s, actual e-mail
% address or url should go in the {}'s for \email and \homepage.
% Please use the appropriate macro foreach each type of information

% \affiliation command applies to all authors since the last
% \affiliation command. The \affiliation command should follow the
% other information
% \affiliation can be followed by \email, \homepage, \thanks as well.
\affiliation{Argonne National Laboratory, Argonne, Illinois 60439} \affiliation{University
of Birmingham, Birmingham, United Kingdom} \affiliation{Brookhaven National Laboratory,
Upton, New York 11973} \affiliation{California Institute of Technology, Pasadena,
California 91125} \affiliation{University of California, Berkeley, California 94720}
\affiliation{University of California, Davis, California 95616} \affiliation{University of
California, Los Angeles, California 90095} \affiliation{Carnegie Mellon University,
Pittsburgh, Pennsylvania 15213} \affiliation{University of Illinois at Chicago, Chicago,
Illinois 60607} \affiliation{Creighton University, Omaha, Nebraska 68178}
\affiliation{Nuclear Physics Institute AS CR, 250 68 \v{R}e\v{z}/Prague, Czech Republic}
\affiliation{Laboratory for High Energy (JINR), Dubna, Russia} \affiliation{Particle
Physics Laboratory (JINR), Dubna, Russia} \affiliation{University of Frankfurt, Frankfurt,
Germany} \affiliation{Institute of Physics, Bhubaneswar 751005, India} \affiliation{Indian
Institute of Technology, Mumbai, India} \affiliation{Indiana University, Bloomington,
Indiana 47408} \affiliation{Institut de Recherches Subatomiques, Strasbourg, France}
\affiliation{University of Jammu, Jammu 180001, India} \affiliation{Kent State University,
Kent, Ohio 44242} \affiliation{Institute of Modern Physics, Lanzhou, China}
\affiliation{Lawrence Berkeley National Laboratory, Berkeley, California 94720}
\affiliation{Massachusetts Institute of Technology, Cambridge, MA 02139-4307}
\affiliation{Max-Planck-Institut f\"ur Physik, Munich, Germany} \affiliation{Michigan
State University, East Lansing, Michigan 48824} \affiliation{Moscow Engineering Physics
Institute, Moscow Russia} \affiliation{City College of New York, New York City, New York
10031} \affiliation{NIKHEF and Utrecht University, Amsterdam, The Netherlands}
\affiliation{Ohio State University, Columbus, Ohio 43210} \affiliation{Panjab University,
Chandigarh 160014, India} \affiliation{Pennsylvania State University, University Park,
Pennsylvania 16802} \affiliation{Institute of High Energy Physics, Protvino, Russia}
\affiliation{Purdue University, West Lafayette, Indiana 47907} \affiliation{Pusan National
University, Pusan, Republic of Korea} \affiliation{University of Rajasthan, Jaipur 302004,
India} \affiliation{Rice University, Houston, Texas 77251} \affiliation{Universidade de
Sao Paulo, Sao Paulo, Brazil} \affiliation{University of Science \& Technology of China,
Hefei 230026, China} \affiliation{Shanghai Institute of Applied Physics, Shanghai 201800,
China} \affiliation{SUBATECH, Nantes, France} \affiliation{Texas A\&M University, College
Station, Texas 77843} \affiliation{University of Texas, Austin, Texas 78712}
\affiliation{Tsinghua University, Beijing 100084, China} \affiliation{Valparaiso
University, Valparaiso, Indiana 46383} \affiliation{Variable Energy Cyclotron Centre,
Kolkata 700064, India} \affiliation{Warsaw University of Technology, Warsaw, Poland}
\affiliation{University of Washington, Seattle, Washington 98195} \affiliation{Wayne State
University, Detroit, Michigan 48201} \affiliation{Institute of Particle Physics, CCNU
(HZNU), Wuhan 430079, China} \affiliation{Yale University, New Haven, Connecticut 06520}
\affiliation{University of Zagreb, Zagreb, HR-10002, Croatia}
\author{B.I.~Abelev}\affiliation{University of Illinois at Chicago, Chicago, Illinois
60607} \author{M.M.~Aggarwal}\affiliation{Panjab University, Chandigarh 160014, India}
\author{Z.~Ahammed}\affiliation{Variable Energy Cyclotron Centre, Kolkata 700064, India}
\author{B.D.~Anderson}\affiliation{Kent State University, Kent, Ohio 44242}
\author{D.~Arkhipkin}\affiliation{Particle Physics Laboratory (JINR), Dubna, Russia}
\author{G.S.~Averichev}\affiliation{Laboratory for High Energy (JINR), Dubna, Russia}
\author{Y.~Bai}\affiliation{NIKHEF and Utrecht University, Amsterdam, The Netherlands}
\author{J.~Balewski}\affiliation{Indiana University, Bloomington, Indiana 47408}
\author{O.~Barannikova}\affiliation{University of Illinois at Chicago, Chicago, Illinois
60607} \author{L.S.~Barnby}\affiliation{University of Birmingham, Birmingham, United
Kingdom}
% ...

\collaboration{STAR Collaboration}\noaffiliation

%Collaboration name if desired (requires use of superscriptaddress
%option in \documentclass). \noaffiliation is required (may also be
%used with the \author command).
%\collaboration can be followed by \email, \homepage, \thanks as well.
%\noaffiliation

\date{\today}

\begin{abstract}
We report the first measurement of the opening angle distribution between pairs of jets
produced in high-energy collisions of transversely polarized protons. The measurement
probes (Sivers) correlations between the transverse spin orientation of a proton and the
transverse momentum directions of its partons.  With both beams polarized, the wide
pseudorapidity ($-1 \leq \eta \leq +2$) coverage for jets permits separation of Sivers
functions for the valence and sea regions. The resulting asymmetries are all consistent
with zero and considerably smaller than Sivers effects observed in semi-inclusive deep
inelastic scattering (SIDIS). We discuss theoretical attempts to reconcile the new results
with the sizable transverse spin effects seen in SIDIS and forward hadron production in pp
collisions.

\end{abstract}

% insert suggested PACS numbers in braces on next line
% insert suggested keywords - APS authors don't need to do this
%\keywords{}

%\maketitle must follow title, authors, abstract, \pacs, and \keywords
\maketitle

\bigskip
%The decomposition of the proton's intrinsic spin among helicity preferences and orbital
%angular momentum of its quark and gluon constituents is a topic of intense interest
%\cite{spin_review} and ongoing experiments
%\cite{PHENIX_pi0_ALL,STAR_jet_ALL,COMPASS_dihadron}. Parton orbital contributions are
%particularly difficult to measure, with one possible manifestation being the so-called
%Sivers effect \cite{Sivers90}: a correlation of parton transverse momentum $({\vec k}_T)$
%with the proton's spin $({\vec s}_{\rm p})$ and momentum $({\vec p}_{\rm p})$ directions,
%yielding $\langle {\vec s}_{\rm p} \cdot ({\vec p}_{\rm p} \times {\vec k}_T) \rangle
%~\neq 0$. Such a three-vector correlation is allowed, without violating time reversal
%invariance, if orbital components of the proton's light-cone wave function combine with
%initial (ISI) and/or final-state interaction (FSI) contributions to the process of
%interest \cite{Brodsky02,Collins03}.  Assuming factorization of perturbative quantum
%chromodynamics (pQCD) cross sections, some ISI/FSI effects can be subsumed in
%gauge-invariant parton distribution (Sivers) functions that depend on both ${\vec k}_T$
%and longitudinal momentum fraction $x_B$.

Hard scattering of light quarks has little sensitivity to one quark's spin orientation
transverse to the scattering plane, due to helicity conservation (chiral symmetry) in the
limit of zero quark mass for both quantum chromodynamics (QCD) and electrodynamics.
Nonetheless, sizable sensitivity to the transverse spin of a \emph{proton} has been
observed at high energies in both semi-inclusive deep inelastic scattering (SIDIS) of
electrons \cite{HERMES} and proton-proton collision processes with cross sections well
described by perturbative QCD (pQCD) \cite{FPDasym}. Theoretical interpretations of these
results \cite{Vo05} attribute them to a combination of soft QCD spin-dependent features of
the proton wave function and of the final-state fragmentation of the struck quark into a
hadron jet. Experiments that can unravel these contributions are essential to understand
high-energy hadron spin dynamics.

Of particular interest, since it arises from orbital contributions to the proton spin
\cite{Brodsky02}, is the Sivers effect \cite{Sivers90}: a correlation ($\langle {\vec
s}_{\rm p} \cdot ({\vec p}_{\rm p} \times {\vec k}_T) \rangle ~\neq 0$) of initial-state
parton transverse momentum $({\vec k}_T)$ with the proton's spin $({\vec s}_{\rm p})$ and
momentum $({\vec p}_{\rm p})$. This three-vector correlation evades time-reversal
violation when orbital components of the proton's light-cone wave function combine with
initial (ISI) and/or final-state interaction (FSI) contributions to the scattering process
\cite{Brodsky02,Collins02}.  In the spirit of pQCD factorization of hadron cross sections,
the Sivers effect involves parton distribution (Sivers) functions that depend on both
${\vec k}_T$ and longitudinal momentum fraction $x_B$. In contrast to ordinary
factorization, gauge invariance demands that Sivers functions incorporate pQCD-calculable,
but process-dependent, ``gauge link factors" describing the partonic ISI/FSI.  These lead
to a predicted sign change between SIDIS and Drell-Yan processes \cite{Collins02,Bo03}.

A non-zero Sivers effect revealed \cite{Hermes05} in SIDIS pion production from a
transversely polarized proton target can be fitted with Sivers functions of opposite sign
and different magnitude for $u$ vs. $d$ quarks \cite{Vo05}. This account can be tested by
treating within a common framework Sivers asymmetries measured for other pQCD processes,
such as jet production in pp collisions \cite{STAR_jet_ALL}. For colliding proton beams
moving along the $\pm {\hat z}$-axis and vertically ($\pm {\hat y}$) polarized, the Sivers
effect gives a preferential sideways ($\pm {\hat x}$) kinematic boost to jet momenta,
causing \cite{Bo04} a spin-dependent average deviation from 180$^\circ$ azimuthal opening
angle between jets from a hard two-body parton scattering.  We report the first
measurement of this di-jet asymmetry, which probes gluon, as well as quark, Sivers
functions. The data were taken in 2006 with $\sqrt{s}=200$ GeV transversely polarized
proton beams at the Relativistic Heavy-Ion Collider (RHIC), providing 1.1 pb$^{-1}$ of
luminosity integrated by the STAR detector \cite{STAR_NIM}.

Continuous operation of two Siberian snakes \cite{Snakes} in each RHIC ring guaranteed
that the beam polarizations were vertical at STAR. The spin orientation alternated for
each successive bunch of one beam and for each pair of bunches of the other. Four distinct
alternation patterns were used for different beam stores to minimize false asymmetries
from accidental correlations of beam properties with bunch number. Beam polarizations,
monitored during each store by proton-carbon Coulomb-nuclear interference polarimeters
\cite{CNI}, averaged 59\% (57\%) for the $+{\hat z}$ ($-{\hat z}$) beam for this analysis,
with statistical uncertainties far smaller than the $\pm 12\%$ relative uncertainty in the
(online) polarimeter calibration.

The detector subsystems critical to the present measurements are the barrel (BEMC) and
endcap (EEMC) electromagnetic calorimeters \cite{STAR_NIM}, with full azimuthal $(\phi)$
coverage spanning pseudorapidities $|\eta| \leq 0.98$ and $1.08 \leq \eta \leq 2.0$,
respectively. The EMC's are subdivided into towers that subtend small regions in $\Delta
\eta$ and $\Delta \phi$. Tower gains are calibrated, to a precision $\approx \pm 5\%$ to
date, with minimum-ionizing particles and electrons tracked with STAR's time projection
chamber (TPC).  Digitized tower signals are summed in STAR trigger hardware over $\Delta
\eta \times \Delta \phi \approx 1.0 \times 1.0$ ``jet patches".  The hardware triggers
used required: (1) a transverse energy sum $E_T > 4.0$ GeV for at least one BEMC or EEMC
jet patch; (2) $E_T^{tot} > 14$ GeV summed over the full EMC; and (3) coincident signals
indicating a valid collision from forward ($3.3 \leq |\eta| \leq 5.0$) beam-beam counters
(BBC) at each end of the STAR detector \cite{BBC}. A software (level 2) trigger then
passed only that subset of events with at least two localized (to $\Delta \eta \times
\Delta \phi = 0.6 \times 0.6$) EMC energy depositions, with $E_{T1(2)} \geq 3.6 ~(3.3)$
GeV and $|\phi_1 - \phi_2| \geq 60^\circ$.

The trigger selectivity for di-jets is illustrated in Fig.\ 1(a-c) by EMC information from
the level 2 processor. The azimuthal angles $\phi_{1,2}$ (referred to the horizontal
$+{\hat x}$-axis in the STAR coordinate frame) and pseudorapidities $\eta_{1,2}$ (measured
with respect to $+{\hat z}$) of the two jet axes are obtained from $E_T$-weighted
centroids of the EMC tower locations in the level 2 jet clusters. The $\eta$ values use an
event vertex determined with coarse resolution ($\sigma_z \approx 30$ cm) from the time
difference between the two BBC's.  The correlation in Fig.\ 1(a) is dominated by intense
di-jet ridges centered around $|\phi_1 - \phi_2| = 180^\circ$.

\begin{figure}
\centering \leavevmode \epsfverbosetrue \epsfclipon \epsfxsize=7.5cm
\epsffile{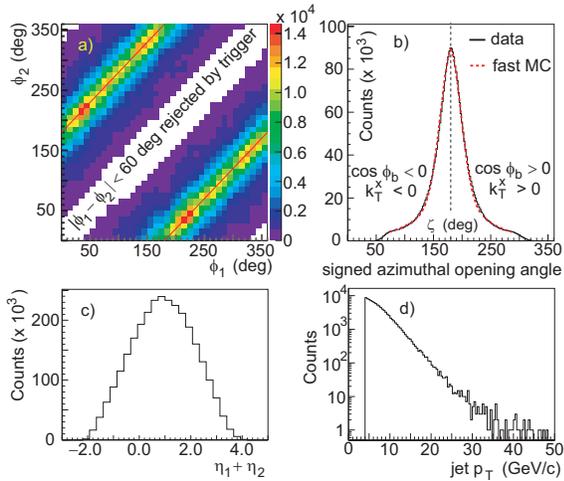} \caption{ Distributions of events that pass the STAR di-jet
trigger with respect to (a) di-jet azimuthal angles, (b) signed azimuthal opening angle
($\zeta$), and (c) pseudorapidity sum, all using EMC energies only. The $p_T$ distribution
(d) is from full jet reconstruction for 2\% of all runs analyzed. The fit in (b) is from
fast Monte-Carlo simulations described in the text.} \label{fig1}
\end{figure}

Initial-state ${\vec k}_T$ is manifested in a given event by a tilt of the jet axes,
characterized by the deviation $|\phi_1 - \phi_2| - 180^\circ$ and the di-jet bisector
angle $\phi_b$.  The Sivers analysis combines these features in a ``signed" azimuthal
opening angle $\zeta$, chosen $> 180^\circ$ when $\cos \phi_b > 0$ (implying $k_T^x > 0$)
and $< 180^\circ$ otherwise. STAR's left-right symmetric di-jet acceptance, reflected in
the $\zeta$ symmetry in Fig.\ 1(b), minimizes systematic errors in our Sivers asymmetries.

The $\zeta$ distribution shape is well reproduced by ``fast" Monte-Carlo (MC) simulations
discussed below. The peak width ($\sigma_{\zeta} \approx 20^\circ$) is dominated by
intrinsic $k_T$ distributions of the scattering partons, with smaller $\phi$ resolution
contributions from the use of EMC energy alone for partial jet reconstruction
($\sigma_\phi^{\rm EMC-full} = 3.9^\circ$), and from deviations between parent parton
directions and even fully reconstructed jet axes ($\sigma_\phi^{\rm full-parton} =
5.0^\circ$). These resolutions were determined, respectively, from the data themselves and
from simulations utilizing the PYTHIA 6.205 event generator \cite{PYTHIA} and GEANT
\cite{GEANT} modeling of the detector response. In the first case we compared, for a small
sample of runs, the $\phi ,~ \zeta$ and $\eta$ values (the latter yielding
$\sigma_\eta^{\rm EMC-full} = 0.07$) determined at trigger level and from full jet
reconstruction including offline gain calibrations and TPC tracks. Full reconstruction,
following the approach in \cite{STAR_jet_ALL}, but with a jet cone radius of 0.6 and $p_T$
threshold of 4.0 GeV, does not greatly improve the net parton directional resolution
($\sigma_\phi^{\rm EMC-parton} = 6.3^\circ$). Thus, the trigger-level di-jet analysis
reported here is sufficient to explore initial results and their implications for
theoretical descriptions of the Sivers effect.

The transverse momentum ($p_T$) distribution from full jet reconstruction (Fig.\ 1(d))
indicates dominance of partons with $x_T \equiv 2p_T/\sqrt{s} \approx 0.05 - 0.10$. The
actual $x_B$ range probed is broad due to the $\eta$ coverage in Fig.\ 1(c). In a
leading-order parton-parton scattering interpretation, $\eta_1 + \eta_2 = \ln
(x_B^{+z}/x_B^{-z})$.  The range $2 < (\eta_1 + \eta_2) < 3$ is then primarily sensitive
to $x_B^{+z} \approx 0.1 - 0.4$, $x_B^{-z} \approx 0.01 - 0.04$, so that the two beams
provide complementary information on valence- and sea-dominated regions.

Fast MC simulations in Fig.\ 2 illustrate some Sivers asymmetry measures. Two-parton
scattering events were generated with a uniform distribution in $\phi_1$ (and $|\phi_2 -
\phi_1| = 180^\circ$) and a $p_T$ distribution reproducing Fig.\ 1(d). Each parton was
given a random initial-state ${\vec k_T}$ drawn from a model distribution centered about
zero for the $y$-component, but about $\pm \langle k_T^x \rangle$ for the $x$-component in
a polarized proton, with the sign correlated with ${\vec s}_{\rm p} \times {\vec p}_{\rm
p}$ to simulate the Sivers effect. The sum ${\vec k}_T^{+z} + {\vec k}_T^{-z}$ was added
to the initially thrown outgoing momenta to deduce boosted azimuthal angles that could
then be further smeared with a Gaussian of $\sigma_\phi^{\rm EMC-parton} = 6.3^\circ$.

\begin{figure}
\centering \leavevmode \epsfverbosetrue \epsfclipon \epsfxsize=7.5cm
\epsffile{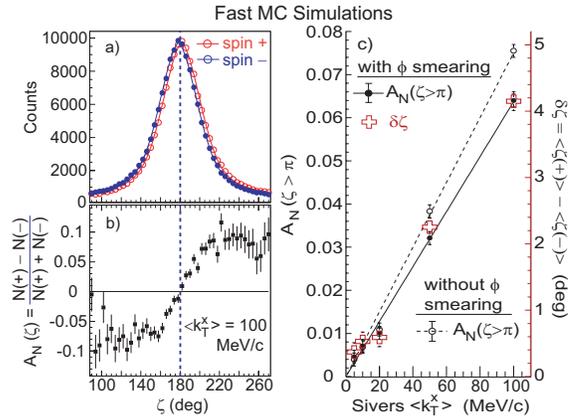} \caption{ Fast MC simulations of di-jet Sivers asymmetries: (a)
the \Z spectra for two beam spins and (b) the resulting yield asymmetry \emph{vs} \Z for
an assumed spin-dependent centroid shift $\pm \langle k_T^x \rangle = \pm$100 MeV/c. (c)
The spin-dependent \Z centroid shift (right scale) and \Z-integrated analyzing power (left
scale) \emph{vs} assumed $\langle k_T^x \rangle$, including (solid line) and excluding
(dashed) $\phi$ resolution smearing of the reconstructed jet. } \label{fig2}
\end{figure}

For Figs.\ 1(b) and 2, the model $k_T$ distribution combines a Gaussian peak with
symmetric exponential tails enhancing larger $|k_T^{x,y}|$, as needed to reproduce the
roughly flat $\zeta$ spectrum wings. Full event reconstruction shows these wings to be
dominated by multi-jet events, reflecting higher-order pQCD processes, where only the two
jets with highest EMC energy were analyzed at level 2. With $\phi$ smearing included, the
$k_T$ distribution fitted to Fig.\ 1(b) has an rms width $\langle (k_T^{x,y})^2 \rangle
^{1/2}$ = 1.26 GeV/c, consistent with the trend of earlier particle correlation results
\cite{ktwidth} from pp collisions. The linear relationship of single-spin observables to
$\langle k_T^x \rangle$ seen in Fig.\ 2(c) is rather insensitive to details of the $k_T$
distribution shape.

Figures 2(a,b) show that the primary Sivers manifestation is a spin-dependent $\zeta$
centroid shift, leading to a spin up \emph{vs.} down yield asymmetry antisymmetric about
$\zeta = 180^\circ$, as predicted in \cite{Bo04}. We sort real data into statistically
independent $\zeta$ distributions for the four beam spin combinations ++, etc., where the
first (second) index is the sign of the ${\hat y}$ polarization component at STAR for the
$+{\hat z}$ ($-{\hat z}$) beam. To compare with predictions integrated over the $k_T$
distributions, we extract analyzing powers $A_N^{\pm z}(\zeta > \pi)$ averaged over
$\zeta$ and $\phi_b$, by fitting asymmetries measured for individual $|\cos \phi_b|$ bins:

\begin{equation}
\label{ANintegrated} f P_{\pm z} |\cos \phi_b| A_N^{\pm z}(\zeta \! > \! \pi) = [r_{\pm
z}(\phi_b) \! - \! 1]/[r_{\pm z}(\phi_b) \! + \! 1],
\end{equation}

\noindent where the cross-ratios $r$ exploit the antisymmetry in Fig.\ 2(b) by treating
di-jet yields $N_{ij}$ with spin-up and $\zeta
> \pi$ as equivalent to spin-down, $\zeta < \pi$. For example:

\begin{equation}
\label{crossratio} r_{+z}(\phi_b) \equiv \sqrt{\frac{\sum {N_{+j}(\zeta \! > \!
\pi,\phi_b)}} {\sum {N_{-j}(\zeta \! > \! \pi,\phi_b)}}\cdot \frac{\sum {N_{-j}(\zeta \! <
\! \pi,\phi_b)}} {\sum {N_{+j}(\zeta \! < \! \pi,\phi_b)}}},
\end{equation}

\noindent  where sums extend over $-{\hat z}$-beam spin states $j \! = \! +,-$. The
cross-ratio eliminates the need for independent relative luminosities for different spin
combinations and cancels several potential systematic errors.  $(P_{\pm z}) |\cos \phi_b
|$ denotes beam polarization components normal to the di-jet bisector within each $|\cos
\phi_b|$ bin. The factor $f = 0.85\pm 0.07$ in Eq.~\ref{ANintegrated} corrects for
dilution of a parton-level asymmetry by $\phi$ resolution smearing (compare solid and
dashed lines in Fig.\ 2(c)), with an uncertainty to allow for model-dependence in
determining $f$ from simulations. The equivalent of Eq.~\ref{crossratio} for $r_{-z}$ has
yields $N_{i-}(\zeta
> \pi)$ (summed over $+{\hat z}$-beam spin states $i$) in the numerator.  This gives
$A_N \! > \! 0$ when ${\vec k}_T$ points preferentially leftward for a spin-up beam,
following the Madison \cite{Madison}, rather than the opposite Trento \cite{Trento}
convention used in \cite{Vo05}.

The measured asymmetries, integrated over $|\zeta - \pi| \leq 68^\circ$, are compared to
calculations \cite{Bo07} in Fig.\ 3. The systematic error bands combine in quadrature the
$f$ uncertainty and the effect of multi-jet contributions to the $\zeta$ distribution
wings. Limits on the latter effect are deduced by looking for variations in $r_{\pm z}$,
beyond statistical fluctuations, when we extract yields alternatively by changing the
$\zeta$ integration range or subtracting a constant baseline fitted to the $\zeta$ wings
independently for each spin state. We neglect much smaller instrumental asymmetries from
bunch-to-bunch variations in beam path or in azimuthally localized beam background.

\begin{figure}
\centering \leavevmode \epsfverbosetrue \epsfclipon \epsfxsize=8.0cm
\epsffile{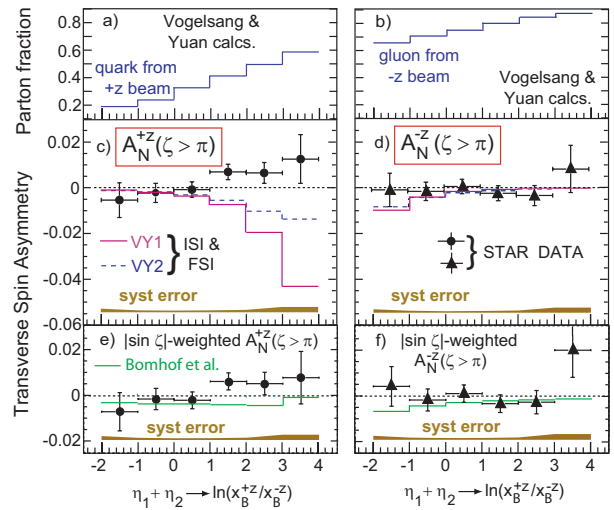} \caption{Measured and calculated asymmetries vs. di-jet
pseudorapidity sum for $+{\hat z}$ (left) and $-{\hat z}$ (right) beams. (a,b): Fraction
of the calculated di-jet cross section with a quark (gluon) from the $+{\hat z} ~(-{\hat
z})$ beam. (c,d): Unweighted asymmetries compared with pQCD calculations \cite{Bo07}
(histograms) for two models of quark Sivers functions fitted to SIDIS results
\cite{Hermes05}.  (e,f): Asymmetries for $|\sin \zeta |$-weighted yields, compared with
calculations \cite{Bo07,footnote2} based on twist-3 quark-gluon correlations. Vertical
(horizontal) bars on the data indicate statistical uncertainties (bin widths). The
systematic error bands exclude a $\pm 12\%$ beam polarization normalization uncertainty. }
\label{fig3}
\end{figure}

The measured asymmetries are consistent with zero, and remain so for higher software EMC
$E_T$ thresholds. BBC yields analyzed with the same code reproduce the associated non-zero
asymmetry \cite{BBC} in both magnitude and sign. Our results are an order of magnitude
smaller than $\pi^+$ SIDIS Sivers asymmetries \cite{Hermes05}, for predominant di-jet
sensitivity (see Fig.\ 3(a-b)) to both high-$x_B$ quarks ($A_N^{+z}(\eta_1 + \eta_2
\gtrsim 2)$) and low-$x_B$ gluons ($A_N^{-z}(\eta_1 + \eta_2 \gtrsim 2)$). The
$\eta$-integrated sample ($2.6 \times 10^6$ di-jet events) has mean $\langle A_N^{\pm
z}(\zeta \! > \! \pi) \rangle = 0$ within statistical uncertainties $\approx \pm 0.002$,
probing (see Fig.\ 2(c)) Sivers $\langle k_T^x \rangle$ preferences as small as $\sim \pm
3$ MeV/c, or $\pm 0.2\%$ of $\langle (k_T^{x,y})^2 \rangle ^{1/2}$.

Recent theory breakthroughs \cite{Bac05,Ji06} and our preliminary results \cite{spin06}
have stimulated rapid evolution in treatments of transverse single-spin asymmetries (SSA).
Bacchetta \emph{et al.} \cite{Bac05} deduced the gauge link structure for hadron or jet
production in pp collisions, where both ISI and FSI contribute, with opposite phases. Ji
\emph{et al.} \cite{Ji06} demonstrated strong overlap between Sivers effects and twist-3
quark-gluon correlations (QGC) \cite{Qiu99}. The pQCD calculations \cite{Bo07} in Fig.\ 3
exploit these developments to incorporate cancellations that were absent or less severe in
predictions \cite{Vo05} made before the measurements. The calculations use one set of
unpolarized distribution functions, yielding the parton contribution fractions in Fig.\
3(a-b), but three different models of $u$- and $d$-quark Sivers functions in Figs.\
3(c-f). All assume zero gluon Sivers function. They are integrated over a $p_T$ range
(5--10 GeV/c) well matched to our data, and further over the STAR $\eta$ acceptance
\cite{Bo07}. We have reversed the sign of the calculated $A_N$ to apply the Madison
convention.

The calculations in Fig.\ 3(c-d) use \cite{Bo07} quark Sivers functions fitted \cite{Vo05}
to SIDIS data \cite{Hermes05} with the $d$-quark functional form tied either to $u(x_B)$
(VY1) or $d(x_B)$ (VY2) unpolarized distribution functions. For $\eta_1 + \eta_2 \gtrsim
2$ the $A_N^{+z}$ predictions reflect the sizable HERMES asymmetries, diluted \cite{Bo07}
by partial $u$ \emph{vs} $d$ and ISI \emph{vs} FSI (the latter were missing in
\cite{Vo05}) cancellations, while $A_N^{-z} \approx 0$ because gluon Sivers effects are
ignored.

Figure 3(e-f) compares $A_N$ measured and calculated \cite{Bo07} with yields in Eq.\ (2)
weighted by $|\sin \zeta|$ \cite{footnote2}, as needed to connect to a more robustly
interpretable gauge link structure \cite{Bac05}, given the apparent breakdown of
factorization for back-to-back dijets \cite{Col07}. The measurements, consistent with zero
at all $\zeta$, are hardly affected by the weighting, but the calculations sample a
different Sivers function moment that can no longer be constrained by unweighted SIDIS
asymmetries. Taking constraints instead from QGC fits \cite{Kou06} to $A_N$ for inclusive
forward hadron production in pp collisions \cite{FPDasym,BrahmsAN} gives di-jet $A_N$
comparable in magnitude to our data, via more complete ISI vs. FSI and $u$ vs. $d$
cancellations \cite{Bo07}. The $u-d$ cancellation can be tested in the future by filtering
quark flavors with the leading hadron's charge sign for each jet.

In summary, we report the first measured spin asymmetries for di-jet production in pp
collisions.  The analysis searches for a spin-dependent sideways tilt of the di-jet axes
sensitive to Sivers correlations between the proton's transverse spin and transverse
momentum preferences of its partons. All measured asymmetries are consistent with zero,
whether dominated by partons in the valence or sea regions. Perturbative QCD calculations
can reconcile these results with sizable SSA observed for forward hadron production in pp
and for semi-inclusive deep inelastic scattering via cancelling contributions from $u$ and
$d$ quarks and from initial- and final-state interactions. These data constrain unified
theoretical accounts for transverse SSA in hard pQCD processes, and their connection to
parton orbital momentum.

We thank the RHIC Operations Group and RCF at BNL, and the NERSC Center at LBNL for their
support. This work was supported in part by the Offices of NP and HEP within the U.S. DOE
Office of Science; the U.S. NSF; a sponsored research grant from Renaissance Technologies
Corporation; the BMBF of Germany; CNRS/IN2P3, RA, RPL, and EMN of France; EPSRC of the
United Kingdom; FAPESP of Brazil; the Russian Ministry of Science and Technology; the
Ministry of Education and the NNSFC of China; IRP and GA of the Czech Republic, FOM of the
Netherlands, DAE, DST, and CSIR of the Government of India; Swiss NSF; the Polish State
Committee for Scientific Research; SRDA of Slovakia, and the Korea Sci. \& Eng.
Foundation.

%%\bibliography{basename of .bib file}

\begin{thebibliography}{20}
\bibitem{footnote2} The $|\sin \zeta |$-weighted asymmetries from \cite{Bo07}
are divided by the measured $\langle |\sin \zeta | \rangle ~=~ 0.33$ in Fig.\ 3(e,f). This
rescaling corrects for a failure to weight the calculated unpolarized cross section, as
needed to compare to measured analyzing powers with unit maximum magnitude.
\end{thebibliography}

\end{document}